\documentclass[aps,prl,showpacs,twocolumn]{revtex4-1}
\usepackage{color}
\usepackage{amsfonts}
\usepackage{amssymb}
\usepackage{amsmath}
\usepackage{epsfig}
\usepackage{graphicx}
\usepackage{enumerate}
\usepackage{multirow}
\usepackage{verbatim}
\usepackage{color}

\begin{document}

\title{Quantum phases of disordered flatband lattice fractional quantum Hall systems}
\author{Shuo Yang}
\affiliation{Condensed Matter Theory Center, Department of Physics,
University of Maryland, College Park, Maryland 20742, USA, \\
and Joint Quantum Institute, University of Maryland, College Park, MD 20742, USA}
\author{Kai Sun}
\affiliation{Condensed Matter Theory Center, Department of Physics,
University of Maryland, College Park, Maryland 20742, USA, \\
and Joint Quantum Institute, University of Maryland, College Park, MD 20742, USA}
\author{S. Das Sarma}
\affiliation{Condensed Matter Theory Center, Department of Physics,
University of Maryland, College Park, Maryland 20742, USA, \\
and Joint Quantum Institute, University of Maryland, College Park, MD 20742, USA}

\begin{abstract}
By numerical exact diagonalization techniques, we obtain the quantum phase diagram of the lattice fractional quantum Hall (FQH) systems in the presence of quenched disorder. By implementing an array of local potential traps representing the disorder, we show that the system undergoes a series of quantum phase transitions as the disorder and/or the interaction is tuned. As the strength of potential traps is increased, the FQH state turns into a compressible liquid, and then into a topologically trivial insulator. We use numerically calculated energy gap, quantum degeneracy, Chern number, entanglement spectrum, and fidelity to identify various quantum phases. The connection to continuum FQH effects is also discussed.
\end{abstract}

\pacs{71.10.Fd, 03.75.Ss, 05.30.Fk, 11.30.Er}
\begin{comment}
71.10.Fd 	Lattice fermion models (Hubbard model, etc.) 
03.75.Ss 	Degenerate Fermi gases 
05.30.Fk 	Fermion systems and electron gas (see also 71.10.-w Theories and models of many-electron systems; see also 67.10.Db Fermion degeneracy in quantum fluids)
11.30.Er 	Charge conjugation, parity, time reversal, and other discrete symmetries 
\end{comment}

\maketitle

{\it Introduction}---
As one of the most fascinating discoveries in physics, the fractional quantum Hall effect (FQHE) has been the central focus of intense research~\cite{Tsui1982,Stormer1999,Nayak.08} over the last three decades. Recently, the interest in this topological state has been further enhanced by the theoretical proposal and the numerical discovery of a new kind of QHE in the absence of an external magnetic field, using lattices with nearly-flat topological bands~\cite{Tang.11, SunK.11, Neupert.11}. Extensive work has established that these novel lattice FQH states are topologically identical to the ordinary continuum FQHE observed in 2D electron systems~\cite{ShengDN.11,Qi.11,Regnault.11,Parameswaran.11,Murthy.11,WangYF.11,Bernevig.11,WangYF.11b,WuYL.11}. This recent theoretical realization of the lattice version of the well-established continuum QHE (and FQHE) has created great excitement in the community because it substantially extends the range of possible topological phases which may exist in nature.

Nearly three decades after its discovery, the topological nature of the FQHE is now reasonably well-understood. However, some key questions with considerable experimental implications remain open. Among these challenges, perhaps the most important is understanding the effect of disorder. As a gapped state, a topologically nontrivial FQH phase is stable against weak disorder. As the strength of disorder increases, the FQH phase is suppressed and the system eventually becomes a trivial insulator. Although some transport signature has been detected in numerical studies (e.g. the closing of the mobility gap~\cite{sheng.03}), the change of topology across this quantum phase transition has not been established theoretically, and the theoretical description for such a topological localization transition is unavailable. Experimentally, however, the disappearance of the topological FQHE with increasing disorder is common knowledge.

In most studies of Anderson localization one increases the random disorder by fixing the strength of the impurities (or the distribution of the strengths) and increasing the number of impurities. In this Letter, we adopt a different disorder model, which fixes the number of the impurities and increases the strength of each impurity potential. This construction provides a new path connecting the gapped FQH state at weak disorder with the topologically trivial insulator at strong disorder. This enables us to investigate the phase transition(s) between these two topologically distinct insulators as well as to clearly distinguish the two phases.

Although this new approach to disorder can be deployed in both lattice and 2D electron gas FQH systems, it is more natural to implement this in lattice systems for numerical studies. By increasing the potential strength of impurity sites in a lattice FQH model, this new approach enables us to construct a complete phase diagram, which contains three different phases: a gapped topological FQH state in the weak disorder limit, a trivial insulator phase at strong disorder, and an intermediate compressible gapless metallic liquid phase. To the best of our knowledge, such a quantum phase diagram has never been theoretically obtained in previous studies of FQHE. Even more remarkably, this construction allows us to examine the nature of these phases and the phase transitions by measuring several different observables, including the energy spectrum, the entanglement spectrum, the Chern number, and the fidelity metric. Some of our findings (e.g., the lack of singularities at the topological transition) offer important guiding principles for the general theoretical understanding of quantum topological phase transition.

\begin{figure}
\includegraphics[width=1.0\columnwidth]{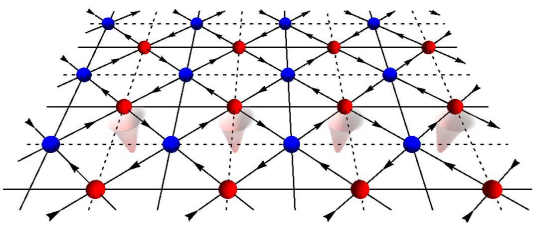}
\caption{(Color online) The checkerboard flatband model with charge impurities marked by the cones beneath the lattice.}
\label{fig:model}
\end{figure}

{\it Model and phase diagram}---
The model we consider is a variation of the topological-flatband model~\cite{SunK.11} on a checkerboard lattice (Fig.~\ref{fig:model}) with the Hamiltonian
\begin{align}
H=H_0+U \sum_{\langle i,j \rangle} n_{i} n_{j} + V \sum_{\langle \langle i,j \rangle \rangle} n_{i} n_{j} +H_{\textrm{impurity}}.
\label{eq:Hamiltonian}
\end{align}
Here, the first three terms are the Hamiltonian of the flatband FQH model proposed in Ref.~\cite{SunK.11,ShengDN.11}, and the last term describes the on-site impurity potential characterizing the disorder strength, which is the tuning parameter for the quantum phase transitions.

The first term ($H_0$) contains all the single-particle hopping kinetic energy terms, 
\begin{align}
H_0= t \sum_{\langle i,j \rangle} e^{i \phi_{ij}} c_{i}^{\dagger} c_{j} + \sum_{\langle \langle i,j \rangle \rangle} t_{ij}' c_{i}^{\dagger} c_{j} +t''\sum_{\langle \langle \langle i,j \rangle \rangle \rangle} c_{i}^{\dagger} c_{j}
+\mathrm{h.c.}\nonumber
\end{align}
Here, $c_{i}$ ($c_{i}^{\dagger}$) is the fermion annihilation (creation) operator at site $i$, while $\left\langle i,j \right \rangle$, $\left\langle \left\langle i,j \right \rangle \right\rangle$, and $\left\langle \left\langle \left\langle i,j \right \rangle \right \rangle \right\rangle$ represent the nearest-neighbor (NN), next-nearest-neighbor (NNN), and next-next-nearest-neighbor (NNNN) bonds, respectively. The $U$ and $V$ terms in Eq.~\eqref{eq:Hamiltonian} are the NN and NNN interactions with $n_{i}=c_{i}^{\dagger}c_{i}$ being the particle number operator. For the hopping strengths, we choose the same parameters as in Ref.~\cite{SunK.11}, i.e., $t=1$, $\phi_{ij} =\pm \pi/4$, $t'=1/(2+\sqrt{2})$ ($t'=-1/(2+\sqrt{2})$) for the solid (dashed) lines, and $t''=1/(2+2 \sqrt{2})$. For a system with $N_{x} \times N_{y}$ unit cells ($2 \times N_{x} \times N_{y}$ sites), we choose $N_{x}N_{y}=3N_{e}$, where $N_{e}$ is the total number of fermions. 
 
The impurities describing the disorder are given by the last term in Eq.~\eqref{eq:Hamiltonian}. Here we select certain lattice sites where local potential traps are introduced to represent the impurities
\begin{align}
H_{\textrm{impurity}}=-\sum_{l} \mu_l n_{l},
\end{align}
where $\mu_l$ gives the strength of the potential trap at the impurity site $l$ and the sum here is taken over the impurity sites only. We make a few non-essential assumptions to simplify the numerical study. We emphasize, however, that these assumptions do not affect our conclusions at all, and we have verified that relaxing them leads to the same qualitative results. We assume that the interaction between the impurity sites and the fermions is attractive ($-\mu_{l}<0$) and that all the impurity sites have the same $\mu$. Due to this assumption ($\mu$ being a constant), in order to achieve an insulating state with a unique ground state in the large $\mu$ limit, we need to require the number of impurity sites $N_i$ to coincide with the number of particles $N_e$. This condition $N_i=N_e$ is non-essential and is not needed if we relax the approximation of $\mu$ being a constant for all impurity sites. In the generic case where $\mu_{l}$ varies from site to site, the large $\mu_{l}$ limit is always a trivial insulating phase, regardless of the value of $N_e$, with all the particles pinned at the $N_e$ sites with  the largest $\mu_{l}$. To further simplify the numerics the impurity sites are chosen along one of the lines in our lattice as shown in Fig.~\ref{fig:model}. 

\begin{figure}
\includegraphics[width=0.95\columnwidth]{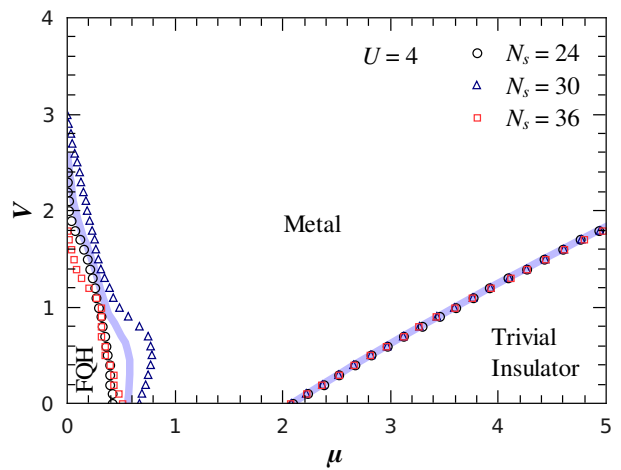}
\caption{(Color online) The phase diagram in the parameter space of impurity
strength ($\mu$) and NNN interaction strength $V$ for different system sizes 
($N_s=2\times 3\times 4$, $2\times 3\times 5$ and $2\times 3\times 6$). 
Three phases are observed, the fractional-quantum-Hall phase (FQH), 
the compressible phase (metal), and the trivial insulator phase. 
The thick blue lines mark the averaged phase boundaries over different system sizes. 
Here we set $U=4$ and $N_e=N_s/6$.}
\label{fig:phasediag}
\end{figure}

Using the exact diagonalization numerical method we establish the phase diagram of this model on finite lattices (Fig.~\ref{fig:phasediag}). The two extreme limits at $\mu=0$ and $\infty$ are easy to understand. For $\mu=0$, Eq.~\eqref{eq:Hamiltonian} recovers the interacting flatband model proposed in Ref.~\cite{SunK.11} (without disorder). At $1/3$  filling, this limit exhibits FQHE when suitable interactions are turned on~\cite{ShengDN.11, Regnault.11}. For very large $\mu$, the fermions are trapped at the impurity sites, resulting in a topologically trivial insulator. For intermediate $\mu$, a gapless compressible phase emerges as an intermediate state, which leads to two phase transitions: (1) a topological transition between the topologically nontrivial FQH phase and the compressible phase, and (2) a typical metal-insulator transition between the compressible liquid and the trivial insulator. We note that for specific value of $U$ and system sizes used in Fig.~\ref{fig:phasediag}, while the transition from the FQH phase to the compressible metal can be driven either by increasing interaction ($V$) or disorder ($\mu$), the transition to the trivial insulator can only happen for large disorder.

\begin{figure}
\includegraphics[width=0.95\columnwidth]{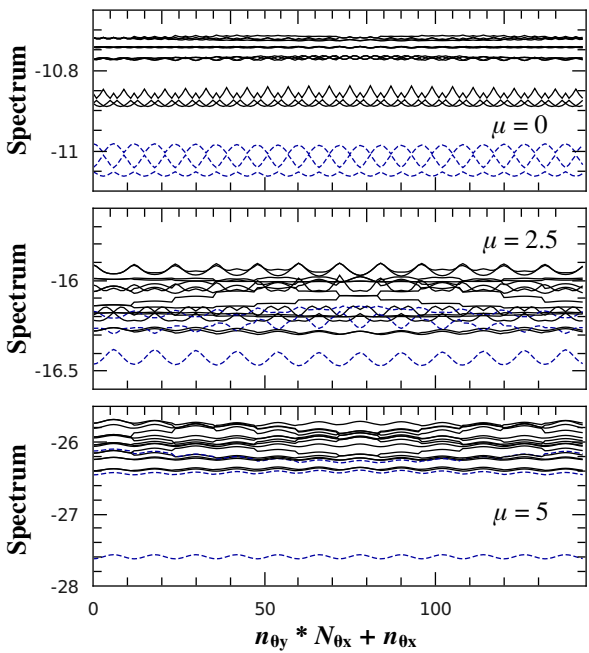}
\caption{(Color online) Energy spectrum as a function of magnetic flux  (a) in the FQH phase with $\mu=0$, (b) in the compressible phase with $\mu=2.5$ and (c) in the trivial insulator phase with $\mu=5$. Here $N_{s}=2 \times 3 \times 6$, $N_{e}=6$, $U=4$, and $V=1$. And we choose
$N_{\theta x}=N_{\theta y}=12$. The blue dashed lines are the three lowest energies of momentum sector $K_{y}=\pi/a$ with $a$ being the lattice constant, and the black solid lines are the energies of the other momentum sectors.}
\label{fig:twistspectrum}
\end{figure}

{\it Numerical measurements}---
The characteristics of these three phases can be seen from the energy spectra upon twisted boundary conditions, which are enforced by requiring all the wave functions to satisfy $T(N_{j}) \left| \Psi \right\rangle=e^{i \theta_{j}} \left| \Psi \right\rangle$, where $j=x$ or $y$ and $T(N_{j})$ is the translation operator along the $j$ direction. Here, we uniformly sample $N_{\theta x}\times N_{\theta y}$ points in the $\theta_{x}$-$\theta_{y}$ plane with $\theta_{j}=2 \pi n_{\theta j}/N_{\theta j}$ and $n_{\theta j}=0,1,\dots,N_{\theta j}-1$. In Fig.~\ref{fig:twistspectrum} the energy spectra of the low energy states are plotted as a function of $n_{\theta y} \times N_{\theta x}+n_{\theta x}$ at different $\mu$. At $\mu=0$ [Fig.~\ref{fig:twistspectrum}(a)], a gapped state with three nearly-degenerate ground states is observed in agreement with Ref.~\cite{ShengDN.11}, which is a characteristic feature of a topological FQH state. For $\mu=2.5$ [Fig.~\ref{fig:twistspectrum}(b)], there is no significant gap, indicating a gapless (compressible) phase. At large $\mu$ [Fig.~\ref{fig:twistspectrum}(c)], we observe a gapped state with a unique ground state (i.e., a trivial non-topological insulator).

\begin{figure}[tbp]
\includegraphics[width=8cm]{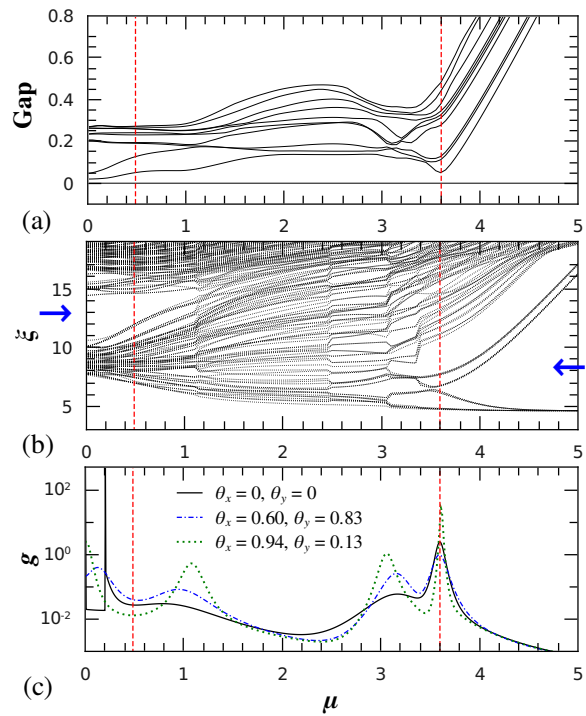}
\caption{Numerical observables as a function of $\mu$. 
(a) Averaged excitation gaps over twisted boundary conditions. 
(b) Particle entanglement spectrum. Here, we
traced out $3$ particles, and the blue arrows mark the gaps 
in the entanglement spectrum. There are $75$ states below the gap
in the FQH state and this number becomes $10$ in the trivial insulator phase, 
in good agreement with the counting of quasi-hole excitations.
(c) Ground-state fidelity metric under periodic and twist boundary conditions,
with $\delta \mu=0.001$.
In all these three panels, the red vertical dashed lines denote the phase boundaries and we 
set $N_{s}=2 \times 3 \times 5$, $N_{e}=5$, $U=4$, and $V=1$.}
\label{fig:gapfidelity}
\end{figure}

To pin down the phase boundaries, we plot the excitation gaps averaged over twisted boundary conditions as a function of $\mu$, $\bar{\Delta}_{i}=\bar{E}_i-\bar{E}_1$, where $i=1$, $2$, $\ldots$, and $\bar{E}_i$ is the average energy of the state $i$. As shown in Fig.~\ref{fig:gapfidelity}(a), the FQH phase at small $\mu$ gives its way to the gapless phase at $\mu\sim 0.5$. As $\mu$ is increased further, an excitation gap reopens at $\mu\sim 3.6$, where the compressible liquid turns into a trivial insulator.

These conclusions are further supported by the entanglement spectrum calculation, $\xi=-2 \mathrm{ln}(\rho_{A})$, which is obtained from the eigenvalues of the reduced density matrix $\rho_{A}=\mathrm{Tr}_{B} \rho$ with $\rho$ being the density matrix of the ground state and $\mathrm{Tr}_B$ representing a partial trace over a part of the system. Here we split the system into two parts, $A$ and $B$, in the particle space. Following the method described in Ref.~\cite{Regnault.11}, we find that at small $\mu$, the entanglement spectrum contains a gap at $\xi\sim 12$, as shown in Fig.~\ref{fig:gapfidelity}(b), and the number of states below this gap agrees with the number of quasi-hole excitations in a FQH state. This gap disappears at $\mu\sim 0.5$ indicating a phase transition from a FQH state into a compressible state. At larger $\mu$ (above $\sim 3.6$), a new gap emerges at $\xi\sim 7$, and the number of states below this gap agrees with the number of excitations in a trivial insulator, suggesting the formation of a non-topological insulator.

It is worthwhile to point out that due to the topological degeneracy, in order to get the correct entanglement spectrum in the FQH phase, it is necessary to use the total density matrix $\rho_{\textrm{FQH}}=\sum_i |\Psi_i\rangle \langle \Psi_i|/3$, which is summed over the three lowest states in the energy spectrum with $i=1$, $2$ and $3$~\cite{Regnault.11}. This is in direct contrast to the entanglement spectrum calculations in other phases where only the lowest state $|\Psi_1\rangle$ is used, $\rho=|\Psi_1\rangle\langle\Psi_1|$. In order to present these two different calculations within the same plot, in Fig.~\ref{fig:gapfidelity}(b), we compute the entanglement spectrum using an approximate density matrix, which is defined as
\begin{align}
\tilde{\rho}=\frac{1}{\mathcal{N}}\sum_{i=1}^3   \exp[-\beta (E_i-E_1)]  |\Psi_i\rangle\langle\Psi_i|.
\end{align}
Here $E_i$ is the energy of the state $i$ and the normalization factor $\mathcal{N}=\sum_{i=1}^{3} \exp[-\beta(E_i-E_1)]$. When $1/\beta$ is much larger than the finite-size gap, but much smaller than the insulating gap, $\tilde{\rho}$ recovers $\rho_{\textrm{FQH}}$ in the FQH phase and $\rho$ in the trivial insulator phase within error bars. Any small difference between $\tilde{\rho}$ and $\rho_{\textrm{FQH}}$ (or  $\rho$) does not change the existence of the gap in the  entanglement spectrum and the number of states below this gap. Therefore, $\tilde{\rho}$ can be used to identify the nature of these phases for all values of $\mu$. 

In addition, we also compute the Chern number in both the insulating phases, topological and trivial, using flux insertion~\cite{Niu.85}. In the FQH phase, because the translational symmetry in our system is broken by impurities, the three nearly-degenerate ground states in our model can no longer be distinguished by different momenta. Therefore, it is not possible to observe the fractional Chern numbers $C=1/3$ in this setup. Instead, the signature of such a FQH state is the total Chern number being unity (similar to the case studied in Ref.~\cite{sheng.03}), which is indeed what we observe here. In the trivial insulator phase, the Chern number is found to vanish as expected. This measurement directly probes the topological structure of the two insulator phases (at small and large $\mu$), and demonstrates clearly the changing of the underlying topology as $\mu$ increases.

{\it Fidelity metric and topological phase transition}---
Finally, we present the ground-state fidelity metric $g$, which has been shown to be a sensitive indicator of quantum phase transitions~\cite{zanardi2006,Gu2010} in certain situations. The fidelity metric $g$ measures the change of the ground state wavefunction in response to a small change of the control parameter, defined as
\begin{align}
g=\frac{2}{N_{s}}\frac{1-F(\mu,\delta \mu)}{(\delta \mu)^{2}}
\end{align}
where $F(\mu,\delta \mu)=\left| \left\langle \Psi(\mu)| \Psi(\mu+\delta \mu) \right\rangle \right|$ is the overlap between the two ground state wavefunctions at $\mu$ and $\mu+\delta \mu$ with $\delta\mu \rightarrow 0$. The value of $g$ remains small inside a quantum phase. However, at a quantum phase transition point, $g$ diverges in the thermodynamic limit, reflecting the singularity associated with the quantum phase transition. In a finite-size system, this divergence is usually regularized by the infrared cutoff resulting in a peak with finite height and width. The existence of such a peak is used as a signature of quantum phase transitions in numerical studies, and this method has been proven to be effective in detecting quantum phase transitions with the only known exception being the Kosterlitz-Thouless transition in 1D quantum systems, where the role of fidelity is still controversial~\cite{Gu2010}. In particular, it has been shown that this method can be used to detect topological transitions in the Kitaev model~\cite{Yang2008} and in the integer quantum Hall effect~\cite{varney2011} as well as in the topological transition between a bosonic FQH phase and a charge-density-wave (CDW) state~\cite{WangYF.11}.

As shown in Fig.~\ref{fig:gapfidelity}(c), a peak at $\mu\sim 3.6$ marks the expected phase transition between the compressible state and the trivial insulator phase. In addition, there are several smaller bumps at smaller values of $\mu$. In some cases, a delta peak may also appear (usually under high-symmetry boundary conditions with $\theta_x$ and $\theta_y$ being $0$ or $\pi$). These bumps and delta peaks do not indicate any phase transitions. Instead, they come from the finite-size effect, and the existence as well as the heights and locations of these bumps (delta peaks) are very sensitive to the boundary conditions ($\theta_x$ and $\theta_y$). In the FQH phase there are three degenerate ground states, whose energies split due to finite-size effects. The values of these splittings (i.e., the finite-size gaps) are very sensitive to microscopic details such as boundary conditions and the value of $\mu$. For certain boundary conditions the finite-size gap between the lowest state and the first excited state may reach a minimum at some $\mu$, which will generate a bump in $g$. In some special cases, these two states may cross at some $\mu$, and this level crossing will induce a delta peak in $g$ as shown in Ref.~\cite{varney2011}. Similar effects also arise in the compressible phase where the (finite-size) gap between the ground state and the first excited state is small. Ignoring these spurious bumps and delta peaks, we find no other feature or signature in $g$ at the topological transition between the FQH state and the compressible liquid. The same is true for the excited states. We therefore conclude that our model reveals a new type of (topological) quantum phase transition which shows no singularity in $g$, in direct contrast to other topological transitions~\cite{Yang2008,varney2011,WangYF.11}. 
 
{\it Conclusion}---
We present a new approach to study the transition from a topological FQH state in weak disorder to a topologically trivial insulator by increasing the strength of disorder. In our calculations, although the locations of the impurity sites are not chosen randomly, the choice of their locations plays no qualitative role. Therefore, we expect that a topologically trivial random insulator obtained via Anderson localization is adiabatically connected to the topologically trivial insulator we find in our model at large $\mu$ although our trivial insulating phase is probably more akin to a pinned quantum charge density wave phase. We emphasize that our approach can, in principle, be deployed in ordinary FQH systems in 2D electron systems. The only difference between the continuum case and the lattice system is that for the continuum model one needs to project all states to the lowest (or some other) Landau level to perform a numerical study, while in lattice models such a projection is unnecessary. In the presence of strong disorder and/or strong interaction, such a projection is not a controlled approximation (for example, due to the strong Landau level coupling), which makes any numerical study in the continuum model less reliable. However, considering the strong similarities between the lattice model and the continuum system, our qualitative phase diagram should in all likelihood apply to the continuum case also.

%%%%%%%%%%
% Acknowledgment%
%%%%%%%%%%
This work is supported by DARPA-QuEST, Microsoft Q, and JQI-NSF-PFC.

%\bibliography{references}
%Merlin.mbs v4.21 2009-07-09.
%

\end{document}